\newcommand{\pjversion}{4.1}
\newcommand{\vdate}{August 1994}
\newcommand{\cernnr}{7420/94}
\newcommand{\PROJT}{\mbox{PROJET}}
\newcommand{\PROJ}{{\tt \PROJT}\ }
\newcommand{\PROJS}{{\tt \PROJT}\ }
\newcommand{\PROJP}{{\tt \PROJT}}

\newcommand{\nonu}{\nonumber\\}
\newcommand{\porder}[1]{\mbox{${\cal O}(#1)$}}

\newcommand{\GeV}{\mbox{GeV}}
\newcommand{\SH}{{S_H}}
\newcommand{\prelim}{($\bowtie$)}
\newcommand{\p}{\tt}
\newcommand{\progline}[1]{\makebox[1.5cm][l]{}#1\\}
\newcommand{\proglineend}[1]{\makebox[1.5cm][l]{}#1}
\newcommand{\commline}[1]{\makebox[1.5cm][l]{c}#1\\}

\newcommand{\fulllineend}[1]{#1}
\newcommand{\pprog}[1]{
           \vspace{0.2cm}
           {\p
           \begin{minipage}{15cm}
           #1
           \end{minipage}
           }
           }
\documentstyle[11pt,fleqn]{article}

\newlength{\dinwidth}
\newlength{\dinmargin}
\begin{document}

\begin{titlepage}

\hfill
\hspace*{\fill}
\begin{minipage}[t]{5cm}
CERN-TH.\cernnr
\end{minipage}

\vspace{1cm}
\begin{center}

{\LARGE
{\p PROJET:}\\
{\rm  Jet Cross Sections\\
in Deeply Inelastic Electron Proton Scattering}\\
\vspace{0.2cm}
{\rm Version \pjversion}\\
\vspace{1cm}
}


\vspace{0.1cm}

{\Large\it
Dirk Graudenz
\footnote{{\em Electronic
mail addresses: graudenz @ cernvm.cern.ch, I02GAU @ DHHDESY3.bitnet}} \\
Theoretical Physics Division, CERN\\
CH--1211 Geneva 23\\
}
\end{center}

\vspace{0.5cm}
\begin{abstract}
\PROJS is a parton level Monte Carlo program for the calculation of jet cross
sections in deeply inelastic electron proton scattering. In its present version
it contains the Born level diagrams for the production of (1+1), (2+1) and
(3+1) jets and the next-to-leading order corrections for the production cross
sections of (1+1) and (2+1) jets for all polarisations of the exchanged virtual
photon. In particular, the full angular correlations between the lepton and jet
momenta are implemented. The program permits the application of acceptance cuts
on all external momenta. For this purpose, the program creates an event record
accessible to the user program with all momenta in the laboratory frame and in
the center of momentum frame of the proton and the virtual photon. This option
is indispensable for phenomenological studies because of the strong dependence
of cross sections on phase space restrictions and the large uncertainty of
fragmentation corrections in the proton direction. Since {\p PROJET} uses the
Monte Carlo integration method for the evaluation of phase space integrals, the
weights of the generated events can be used to produce distributions of
observables related to jet momenta.
\end{abstract}

\vfill
\noindent
\begin{minipage}[t]{5cm}
CERN-TH.\cernnr\\
\vdate
\end{minipage}
\vspace{1cm}
\end{titlepage}


\begin{center}
{\large PROGRAM SUMMARY}
\end{center}

\noindent
{\em Title of the program:} {\p PROJET} version \pjversion.

\noindent
{\em Computer:} Any computer with {\p FORTRAN 77} compiler.

\noindent
{\em Programming language used:} {\p FORTRAN 77}.

\noindent
{\em High speed storage required:} Size of executable program is approximately
2.5 MBytes.

\noindent
{\em No. of cards in combined program and testdeck:} about 40.000.

\noindent
{\em Other programs used:} {\p VEGAS} \cite{Lep78,Lep80}
(multidimensional Monte
Carlo integration routine), {\p PAKPDF} \cite{Cha92},
{\p PDFLIB} \cite{Plo93} (parametrizations of parton
densities).

\noindent
{\em Keywords:} Quantum Chromodynamics (QCD),
Jet Physics, Deeply Inelastic Electron Proton Scattering,
Monte Carlo Simulation.

\noindent
{\em Nature of physical problem:}
In deeply inelastic scattering, the hadronic final state can be analysed by a
jet cluster algorithm resulting in cross sections for the production of (n+1)
jets (including the target remnant jet). It is possible to extract physical
parameters ($\Lambda_{\mbox{QCD}}$, parton densities
$f_i\left(\xi,\mu_f^2\right)$) from
experimentally measured
jet cross sections.

\noindent
{\em Method of solution:}
{\p PROJET} makes it possible to
study these cross sections from a
theoretical point of view by performing the integration of
the differential cross sections over phase space regions that are related to
a certain number of jets. QCD corrections
are included by analytical formulae.

\noindent
{\em Restrictions on the complexity of the problem:}
The program contains jet cross sections for the production
of (1+1), (2+1) and (3+1) jets. QCD corrections
to (1+1) and (2+1) jet cross sections are included for
all polarisations of the exchanged virtual photon.

\noindent
{\em Typical running time:} Depends strongly on the process under
consideration;
the generation of the weight for 1 event takes typically 0.001 to 0.02 seconds
on a DEC or hp workstation, depending on the process; therefore an
integration using 200.000 events typically needs 3 minutes to 1 hour of
CPU time.

\newpage

\section{Introduction}
\label{intro}

Recent results from HERA show evidence for events
with a pronounced jet-like structure in the hadronic final state
\cite{HERA}.
In order to perform quantitative tests of perturbative QCD,
next-to-leading (NLO) order predictions have to be confronted
with experimental data.
In principle, there are two basic tools
to study jet physics from a theoretical point of view:
\begin{itemize}
\item Event generators
\item Parton level Monte Carlo programs.
\end{itemize}

\noindent
Programs of the first class (event generators) simulate the underlying physics
by generating full events with a hadronic final state with probabilites given
by
some hard processes based on exact matrix elements, a leading logarithmic
correction using parton showers and a final hadronisation step based on a
specific model motivated by phenomenology. A certain set of parameters can be
adjusted to fit experimental data very precisely. Event generators are a very
convenient tool to make detector studies. Examples are {\p ARIADNE},
{\p HERWIG},
{\p LEPTO} and {\p PYTHIA} \cite{HERA91}.

\noindent
The second class of programs (parton level Monte Carlo programs) is
based on a different
approach. Their goal is to make predictions using only a very small number of
parameters (basically $\Lambda_{\mbox{QCD}}$
and a certain set of parton densities
$f_i\left(\xi,\mu_f^2\right)$).
The final state consists of partons
clustered by a means of a suitable jet algorithm
and integrated over singular phase space regions.
Therefore assumptions
have to be made on how to map the results to a realistic experimental
situation.
The correspondence between the parton level and the hadron level is made by a
suitable jet definition (in principle
it is assumed that a hard parton accompanied
by collinear and soft partons can be identified with a hadronic jet).
Jet definitions in deeply inelastic scattering
based on a cluster algorithm are described elsewhere
\cite{BKMS89,GM92,Gra90,Gra91,Gra94a,KMS89}, so
no further comments are made on this problem here.
Because of the small number of parameters that are involved
in parton level Monte Carlo programs it it possible to
extract fundamental parameters of the underlying field theory
($\Lambda_{\mbox{QCD}}$and possibly the parton
densities $f_i\left(\xi,\mu_f^2\right)$).

\noindent
\PROJS is a parton level Monte Carlo program for the calculation of jet cross
sections in deeply inelastic electron proton scattering on the basis
of Born terms and NLO corrections.
Recently, another program ({\p DISJET}) for jet cross section calculations
has been published \cite{BM94}.
It can be used to calculate the total cross section
in NLO for (1+1) and (2+1) jet production.
For this purpose, the metric and longitudinal polarisations
of the exchanged virtual photon
as implemented in {\p DISJET} are
sufficient\footnote{For a discussion, see Section~\ref{ME}.}.
In the case of cuts on the jet momenta this statement is no longer
true and a full implementation of the matrix elements is indispensable.
Moreover, jet cross sections
strongly depend on restrictions of the phase space of
outgoing particles. These restrictions appear naturally
in experimental situations because of acceptance cuts for
detectors (an example being given by angle cuts in the case
of the beam pipe at the eP collider HERA) and possibly theoretically
in order to restrict calculations to phase space regions where
the use of fixed order perturbation theory is justified.
In addition,
fragmentation corrections
are not well known in the case of jets in the forward
direction\footnote{The various QCD inspired
models predict different shapes of the transverse energy flow
in the forward direction,
for an overview see \cite{Fel94}.
The forward direction is defined to be the
direction of the incoming proton.}.
In order to enable studies of the kind just mentioned,
{\p PROJET} supplies the user with an event record for
every generated event, thus enabling him to restrict the phase space
arbitrarily.
In addition, since the weight of every generated event
is known as well, distributions and expectation values
of observables depending on external momenta can be studied.
A detailed phenomenological study is in preparation \cite{Gra94b}.

\noindent
It should be stressed that \PROJS is not intended to be a tool that works
like an event generator. First of all, the program does not produce events with
unit weight, but simply uses the Monte Carlo method to do a multidimensional
integration, and secondly, the final state consists of jets and not of hadrons.
Therefore the output of the program can only be compared to experimental
results if the latter are corrected for detector acceptance and processed by
the same jet cluster algorithm (mJADE in the present implementation,
see \cite{GM92}) used in the program.
This means in particular that the jet cut $c$ has to be the analysis
cut of the experimental cluster algorithm. Generating {\p PROJET}
events with a
very small cut and performing a subsequent clustering
is {\em not} a viable procedure from the theoretical point of view
since this does not take QCD corrections into account in the correct way.

\noindent
This manual is organised as follows. In the next section the matrix
elements used in \PROJ are discussed. In Section~\ref{PS}
the general structure of \PROJ is described. The program calculates
the four-vectors of the external particles and stores their values in an event
record. The information contained in the event record is described in
Section~\ref{TER}. A user interface is explained in Section~\ref{UdS}.
Finally a complete list of the program parameters that are implemented
and a list of possible error messages of the program is given
in Section \ref{PP}.

\section{Matrix Elements}
\label{ME}

\noindent
\PROJS integrates matrix elements for the processes
\begin{equation}
\mbox{e}^-\,+\,\mbox{proton} \,\rightarrow\,
\mbox{proton remnant}\,+\,\mbox{n jets}
\end{equation}
for n up to 3. Since the proton remnant is counted as a jet, matrix elements
for (1+1), (2+1) and (3+1) jet production are implemented.

\noindent
The cross section in the (2+1) jet case
can be decomposed as a sum over helicity cross sections
\begin{eqnarray}
\sigma &=&\frac{1+(1-y)^2}{2\,y^2}\,\,\sigma_M
      +\frac{1+(1-y)^2+4(1-y)}{2\,y^2}\,\,\sigma_L\nonu
       &+&\frac{2-y}{y^2}\sqrt{1-y}\,\cos\Phi\,\,\sigma_{\Phi}
       +2\frac{1-y}{y^2}\,\cos2\Phi\,\,\sigma_{2\Phi}
\end{eqnarray}
including those which are mediated by the exchange of a virtual photon
with metric
and longitudinal
polarisation ($\sigma_M$ and $\sigma_L$, respectively)\footnote{Metric
contributions are defined by the contraction
of the hadronic tensor $H_{\mu\nu}$ with the metric
$-g^{\mu\nu}$, and longitudinal contributions by the contraction
with $p_0^\mu p_0^\nu$, where $p_0$ is the momentum of the incoming parton.
Projection operators for the other helicity cross sections can be defined
as well.}
(a detailed discussion
can be found in \cite{KMS89}).
Here and in the following the relevant kinematical
variables are
\begin{eqnarray}
&&x_B=\frac{Q^2}{2Pq},\quad
y=\frac{Pq}{Pk},\quad\nonu
&&\SH=(P+k)^2,\quad
Q^2=-q^2=\SH\,x_B\,y,\quad
W^2=\SH\,(1-x_B)\,y,
\end{eqnarray}
where $P$ is the proton momentum,
$q$ is the momentum of the exchanged virtual photon,
and $k$ is the momentum of the incoming lepton.
$\Phi$ is the angle of the plane defined by
the incoming and outgoing leptons
and that of one outgoing jet and the incoming proton, respectiveley,
in the center of momentum (CM)
frame of the incoming proton and the virtual photon.
Integrated over $\Phi$, the total cross
section can be expressed in terms of the metric and longitudinal cross
sections ($\sigma_M$ and $\sigma_L$)
alone (where in many cases the metric contributions dominate).
This statement is no longer true if one imposes angular cuts on the outgoing
particles, because then the integration domain is restricted, and the
additional
terms do not integrate to zero any longer.

\noindent
For the Born level, the relevant Feynman diagrams and references to the
literature can be found in \cite{BKMS89,GM92,KMS89}.
The Born matrix
elements
have been implemented
in two different ways. One implementation ({\p NPARTONS}~$>$0,
{\p NPARTONS} is a parameter which is explained later on)
uses the contraction of the complete lepton tensor with the hadron tensor, the
other implementation ({\p NPARTONS}~$<$0) uses the contributions
{}from the explicit projections on the various polarisations
of the virtual photon.
Therefore,
longitudinal contributions can be studied separately.

\noindent
The NLO contributions in \PROJS for the production of
(2+1) jets for the metric polarisation of the virtual
photon are those from
\cite{Gra90,Gra91,Gra94a}\footnote{Please note that there
is a sign error in the virtual corrections in \cite{Gra90,BK91}.
This has been corrected in \cite{Gra94a}. All {\p PROJET} versions
since early 1993 use the correct matrix elements.}.
The NLO corrections for the other polarisations of the
exchanged virtual boson can be found in [BK91].
However, they have been recalculated for the present implementation
in order to have a form of the analytical formulas
that could be easily implemented in one of the recent versions of
{\p PROJET}.
These matrix elements
are correct up to terms
\porder{c\log^2c}, where $c$ is the jet cut.
In addition, the NLO contribution for the production
of (1+1) jets (metric and longitudinal polarisation of the virtual photon)
is implemented by subtracting the
analytically integrated (2+1) jet cross section of the order \porder{\alpha_s}
{}from the
total cross section of \porder{\alpha_s}
(see, for example, \cite{AEM79,KMS89,Gra94a}).

\noindent
A complete list of cross sections implemented in \PROJS
can be found in table~\ref{tab1} (see Section~\ref{PP}).

\section{Program Structure}
\label{PS}

\PROJS consists of a set of subroutines. The user has to supply a main program
that controls \PROJS by calls to the subroutine
{\p setpar(ipar,\ value,\ ierr)}.
{\p ipar} specifies whether a parameter of \PROJS has
to be changed or a specific action should be performed. {\p value} is the new
value of the selected parameters (double precision), and {\p ierr}
is an error code.
An error condition is given by {\p ierr}~$\neq$~0.
The possible values of ipar are given in the following list:
\begin{itemize}
\item { {\p ipar=}\makebox[1.5cm][l]{\p (-3):}}\\
Initialize \PROJS and preset all parameters.
\item { {\p ipar=}\makebox[1.5cm][l]{\p (-2):}}\\
Initialize the adaptive integration routine {\p VEGAS}.
\item { {\p ipar=}\makebox[1.5cm][l]{\p (-1):}}\\
Initialize phase space integration. Start integration.
\item { {\p ipar=}\makebox[1.5cm][l]{\p 0:}}\\
No operation (``comment'').
\item { {\p ipar}$>=${\p 1:}}\\
Set a \PROJS parameter. See Section~\ref{PP}.
\end{itemize}

\noindent
If a reference to a parameter could not be resolved (i.e., if \PROJS
does not know this parameter), the user supplied subroutine {\p userset}
is called.

\noindent
The general structure of the main program should be the following:
\begin{itemize}
\item (a) initialize \PROJS and {\p VEGAS} by subsequent calls \,
{\p call setpar(-3,\ 0.d0,\ ierr)}, {\p call setpar(-2,\ 0.d0,\ ierr)}.
\item (b) modify the default parameters of \PROJS by calls
{\p call setpar(ipar,\ value,\ ierr)}.
\item (c) start the integration by
{\p call setpar(-1,\ 0.d0,\ ierr)}.
\item (d) if necessary, repeat steps (b) and (c).
\end{itemize}

\noindent
The user has to supply two subroutines {\p user}, {\p userset}
and a function {\p iusercut}. They are explained in Section~\ref{UdS}.

\noindent
There is an example program {\p MCEX.F} reading a parameter file
{\p PAREX} that calculates various jet cross sections. This program may be
used as a guideline for main programs. To facilitate a check of the program,
the
file {\p RESEX} contains the output from file {\p unit=6}, and the file
{\p PLOTEX} the output from file {\p unit=55}. In this
particular test run program,
the numbers should not be taken
literally, since the integration uses only a small number of points
in order to facilitate running the program, and results
in general differ because the floating point accuracy depends on the machine
where the program is running.

\noindent
\PROJS prints status information on the standard output file ({\p unit=6}).
There is no input required from the standard input file ({\p unit=5}).
All floating point variables in \PROJS are {\p double precision}
variables. The file units {\p unit=55,63,64,65,91} are reserved
as output files for \PROJP.

\noindent
\PROJS uses the following programs:
\begin{itemize}
\item The phase space integrations
in \PROJS are performed by the adaptive integration
routine {\p VEGAS} \cite{Lep78,Lep80} (for (3+1) jets the total phase space
is 8-dimensional).
\item The parton density parametrisations {\p PAKPDF} are from
\cite{Cha92}. In addition, an interface to the {\p PDFLIB} library
\cite{Plo93}
of parton densities is implemented.
\end{itemize}

\noindent
In its present form \PROJ\
contains code that was used for other purposes
(interface to an event generator, histograms, ...). These parts will be removed
in the future when they are no longer needed and should therefore not be used.

\section{The Event Record}
\label{TER}

\noindent
If a generated event fulfills all cuts its momentum four-vectors
in the laboratory frame (LAB frame) and in the CM frame of the
proton and the virtual photon (PVP frame) are stored in
the event
record.
The corresponding variables can be
found in the common blocks {\p /evtrecord/}
and {\p /pvprecord/}.
Additional information is stored in {\p /userwgt/}:

\begin{verbatim}
      integer nmaxent
      parameter (nmaxent=20)
      double precision pout,ppol
      integer nentry,iptype,iident,inout
      common /evtrecord/
     &       pout(nmaxent,4),
     &       ppol(nmaxent,4),
     &       nentry,iptype(nmaxent),iident(nmaxent),inout(nmaxent)

      double precision pvp,pvppol
      common /pvprecord/
     &       pvp(nmaxent,4),
     &       pvppol(nmaxent,4)

      double precision evwgt,evxsect,everror
      integer ievaccpt
      common /userwgt/
     &       evwgt,
     &       evxsect,everror,
     &       ievaccpt
\end{verbatim}

\noindent
The user has access to these common
blocks in the user routine {\p user} and in the function
{\p iusercut}. The number of entries
in the event record is given by {\p
nentry}. For every entry {\p i}
(1 $\leq$ {\p i} $\leq$ {\p nentry}),
{\p pout(i,1..3)} is the 3-momentum of the particle
in the LAB frame (1 is the $x$-direction
which is always in the plane of the incoming electron and the outgoing
electron,
2 is the $y$-direction and 3 is the positive
$z$-direction which is defined to be the direction of the incoming proton;
if the proton is at rest ({\p EPEE1}~$<$~0,
see Section \ref{PP}), then the $z$-direction is defined
to be the direction opposite to the incoming electron)
and {\p pout(i,4)} is its energy (in units of \GeV).
Furthermore, the polar coordinates in the LAB frame
are stored in the array {\p ppol}, namely,
{\p ppol(i,1)} is the energy,
{\p ppol(i,2)} is the modulus of the momentum,
{\p ppol(i,3)} is the polar angle in radians, and
{\p ppol(i,4)} is the azimuthal angle in radians.
The corresponding four momenta in the PVP frame are
stored in the variables {\p pvp} and {\p pvppol},
respectively.

\noindent
The momenta of the following particles are stored in the event record:
\begin{itemize}
\item {\p iident(i)=}\makebox[1.5cm][l]{\p (-4):} incoming proton,
\item {\p iident(i)=}\makebox[1.5cm][l]{\p (-3):} incoming electron,
\item {\p iident(i)=}\makebox[1.5cm][l]{\p (-2):} outgoing electron,
\item {\p iident(i)=}\makebox[1.5cm][l]{\p (-1):} exchanged virtual photon,
\item {\p iident(i)=}\makebox[1.5cm][l]{\p 0:} remnant jet,
\item {\p iident(i)=}\makebox[1.5cm][l]{\p 1,2,...:} parton jets.
\end{itemize}

\noindent
The array {\p iident} allows the identification of the entries in the event
record. In addition, there is an array {\p iptype} that specifies the particle
type:
\begin{itemize}
\item {\p iptype(i)=}\makebox[1.5cm][l]{\p 0:} unspecified particle,
\item {\p iptype(i)=}\makebox[1.5cm][l]{\p 1:} electron,
\item {\p iptype(i)=}\makebox[1.5cm][l]{\p 2:} jet,
\item {\p iptype(i)=}\makebox[1.5cm][l]{\p 3:} proton,
\item {\p iptype(i)=}\makebox[1.5cm][l]{\p 4:} virtual photon.
\end{itemize}

\noindent
Finally, there is an array {\p inout} specifying whether a particle in the
event
record is incoming, internal, or outgoing:
\begin{itemize}
\item {\p inout(i)=}\makebox[1.5cm][l]{\p (-1):} incoming,
\item {\p inout(i)=}\makebox[1.5cm][l]{\p 0:} internal,
\item {\p inout(i)=}\makebox[1.5cm][l]{\p 1:} outgoing.
\end{itemize}

\noindent
The following table is an example for an event record of a (3+1) jet
event with $E_P$~=~820~\GeV, $E_e$~=~26.7~\GeV.
The first column gives the index of the entry, the second column
is {\p iptype(i)}, the third column is {\p iident(i)}, the fourth
column is {\p inout(i)}, and the other columns are the {\p pout(i,j)}
in the order $E$, $p_x$, $p_y$, $p_z$.

\begin{verbatim}
   1   3  -4  -1    0.82000D+03    0.00000D+00    0.00000D+00    0.82000D+03
   2   1  -3  -1    0.26700D+02    0.00000D+00    0.00000D+00   -0.26700D+02
   3   1  -2   1    0.27809D+02    0.11892D+02    0.00000D+00   -0.25138D+02
   4   4  -1   0   -0.11096D+01   -0.11892D+02    0.00000D+00   -0.15613D+01
   5   2   0   1    0.66116D+03    0.00000D+00    0.00000D+00    0.66116D+03
   6   2   1   1    0.71842D+02   -0.53358D+01   -0.10878D+00    0.71643D+02
   7   2   2   1    0.21690D+02   -0.15676D+01   -0.24633D+00    0.21632D+02
   8   2   3   1    0.64201D+02   -0.49885D+01    0.35510D+00    0.64005D+02
\end{verbatim}

\noindent
To obtain the index {\p i} of a particle identification {\p iident1}, there
is a function {\p ievid(iident1)} that returns the index {\p i}.
The following
example stores the momentum four-vector of the outgoing electron in the
variables {\p p0}, {\p p1}, {\p p2}, {\p p3}:

\pprog{
   \noindent
   \commline{identification of the outgoing electron is -2}
   \progline{iident1=-2}
   \commline{get index}
   \progline{i=ievid(iident1)}
   \commline{get four-vector}
   \progline{p0=pout(i,4)}
   \progline{p1=pout(i,1)}
   \progline{p2=pout(i,2)}
   \proglineend{p3=pout(i,3)}
}
\smallskip

\noindent
The relative weight of every event is given by the variable {\p
evwgt}. The sum of {\p evwgt}
of all events in the last integration of {\p VEGAS} is
the cross section for the
process in units of [pb]. Finally, there is a flag {\p ievaccpt}
showing whether an event passed all cuts and the event has been generated
in the final {\p VEGAS} integration
({\p ievaccpt=1}) or the event has been
rejected or has been generated in one of the grid definition iterations of
{\p VEGAS}
({\p ievaccpt=0}).

\noindent
After the integration has been performed,
the integrated cross section is stored in {\p evxsect}, and the estimated
error of the integration in {\p everror} (this error estimate by {\p VEGAS}
is not always reliable, see \cite{Lep78,Lep80}).

\section{User Defined Subroutines}
\label{UdS}

\noindent
The user has to define subroutines

\pprog{
   \noindent
   \progline{subroutine user(iopt,ierr)}
   \progline{integer iopt,ierr}
   \progline{}
   \progline{subroutine userset(ipar,idata,xdata,ierr)}
   \progline{integer ipar,idata,ierr}
   \proglineend{double precision xdata}
}

\noindent
and a function

 \pprog{
   \noindent
   \progline{function iusercut}
   \proglineend{integer iusercut}
}

\noindent
of integer type.

\noindent
{\p user} is called at different stages of the program:
\begin{itemize}
\item {\p iopt=1:} Called after the preset of all parameters (by a call of
{\p setpar(-3,...)}). Here the user can modify values of the presets
for all parameters and open files used for output purposes.
\item {\p iopt=2:} Called after the initialisation of the integration routine
(by a call of {\p setpar(-2,...)}).
\item {\p iopt=3:} Called before the start of the integration (by a call
of {\p setpar(-1,...)}). At this point the user should
initialize histograms and variables used for the averaging of observables.
\item {\p iopt=4:} Called in the case of every event. Here the user
can bin observables into histograms
or update averages. This subroutine is called even if an event has been
rejected
(if it did not satisfy any of the cuts). The variable {\p evwgt} contains the
weight of the event (summing to the total cross section in [pb]). Here the flag
{\p ievaccpt} can be tested.
{\p user(4,...)} is called also
during the grid definition runs of {\p VEGAS}.
\item {\p iopt=5:} Called before
the program terminates. This is the place to write
histogram files to an external file or to print results.
\item {\p iopt=6:} Called in case of an error.
If something went wrong and the program did not crash
before, {\p ierr} contains an error code. If {\p ierr}~$<$~0, the error is
fatal
and \PROJS will stop. If {\p ierr}~$>$~0, the error is not fatal,
and \PROJS will continue.
\end{itemize}

\noindent
{\p userset} is called if a parameter is not known to \PROJP.
{\p ipar} is the number of the requested parameter, {\p idata} is its
integer value and {\p xdata} its double precision value. {\p userset}
must return an error code in {\p ierr}. If the error code is not zero,
\PROJS will issue an error message. {\p userset} will be called once after
{\p setpar(-3,...)} with {\p ipar=-3} to give the user the opportunity to
initialize his own parameters.

\noindent
{\p iusercut} is called after the kinematical variables have been calculated,
but before the cross section is evaluated.
The user can inspect the event record
and impose additional cuts. If an event is to be rejected, {\p iusercut} should
return 0, else 1.

\section{Program Parameters}
\label{PP}

In this section the input parameters that can be used with
{\p setpar} are described in detail.

\noindent
In the following list {\p (VARIABLENAME)} gives the name of the variable
that is affected. It is not necessarily the case that {\p par} is assigned to
the variable, sometimes a function of {\p par} is assigned.
{\p (FLOAT)} means that a parameter takes floating point
values, and {\p (INT)}
means that it takes integer values. {\p (DEF=XXX)} indicates that the
default if no input is given is {\p XXX}.
Parameter options marked with \prelim {} have been implemented due to
requests from {\p PROJET} users.
They are included here for completeness only.
It is recommended to consult the program author or to study the
source code of {\p PROJET} before use is made of one of these parameters
because they are sometimes restricted in use.

\begin{itemize}
\item {\p {\p ipar}=\makebox[1.5cm][l]{2001:}
                        \makebox[3cm][l]{({\p KPAR})}
                        \makebox[3cm][l]{(INT)}
                        \makebox[3cm][l]{(DEF=4)}}\\
Specifies which lepton
variables are fixed and which are integration variables.\\
\makebox[0.7cm][l]{0:} $x_B$ and $y$ fixed;\\
\makebox[0.7cm][l]{2:} $x_B$ and $Q$ fixed;\\
\makebox[0.7cm][l]{4:} integrate over $x_B$ and $Q^2$ within the
cuts \#2030--\#2037.\\
If $x_B$ and $y$ or $Q$ are fixed ({\p KPAR=0,2}),
the program returns $d\sigma/(dx_Bdy)$
in [pb], if {\p KPAR=4}, the integrated cross section in [pb] is returned.
\item {\p {\p ipar}=\makebox[1.5cm][l]{2002:}
                        \makebox[3cm][l]{({\p SH})}
                        \makebox[3cm][l]{(FLOAT)}
                        \makebox[3cm][l]{(DEF=295.93)}}\\
Specifies the CM-energy $\sqrt{\SH}$ of the incoming electron and proton.
\item {\p {\p ipar}=\makebox[1.5cm][l]{2003:}
                        \makebox[3cm][l]{({\p XH})}
                        \makebox[3cm][l]{(FLOAT)}
                        \makebox[3cm][l]{(DEF=0.1)}}\\
Specifies Bjorken-x $x_B$ in the case of {\p KPAR}=0, 2.
\item {\p {\p ipar}=\makebox[1.5cm][l]{2004:}
                        \makebox[3cm][l]{({\p Y})}
                        \makebox[3cm][l]{(FLOAT)}
                        \makebox[3cm][l]{(DEF=0.2)}}\\
Specifies the lepton variable $y$ in the case of {\p KPAR}=0.
\item {\p {\p ipar}=\makebox[1.5cm][l]{2007:}
                        \makebox[3cm][l]{({\p QQ2})}
                        \makebox[3cm][l]{(FLOAT)}
                        \makebox[3cm][l]{(DEF=100.0)}}\\
Specifies the momentum transfer of the photon $\sqrt{Q^2}$
in the case of {\p KPAR}=2.
\item {\p {\p ipar}=\makebox[1.5cm][l]{2008:}
                        \makebox[3cm][l]{({\p CUT})}
                        \makebox[3cm][l]{(FLOAT)}
                        \makebox[3cm][l]{(DEF=0.01)}}\\
Specifies the invariant jet cut $c$. The cut condition
depends on \#2046.
\item {\p {\p ipar}=\makebox[1.5cm][l]{2014:}
                        \makebox[3cm][l]{({\p XIIMIN})}
                        \makebox[3cm][l]{(FLOAT)}
                        \makebox[3cm][l]{(DEF=0.0001)}}\\
Lower bound on the momentum fraction of the incoming parton, its value depends
on the parton densities. It is important to specify this parameter since
otherwise \PROJS could try to evaluate the parton densities in a region where
they are no longer defined, which would result in an error. {\p XIIMIN} is
typically of the order of 0.0001 to 0.00001.
\item {\p {\p ipar}=\makebox[1.5cm][l]{2015:}
                        \makebox[3cm][l]{({\p IPD})}
                        \makebox[3cm][l]{(INT)}
                        \makebox[3cm][l]{(DEF=6)}}\\
Selects the parton density distribution. If a NLO order contribution
of the matrix elements is
included, the parton density parametrization should be using the
$\overline{\mbox{MS}}$-scheme.
\item {\p {\p ipar}=\makebox[1.5cm][l]{2016:}
                        \makebox[3cm][l]{({\p IPDSET})}
                        \makebox[3cm][l]{(INT)}
                        \makebox[3cm][l]{(DEF=2)}}\\
Selects the subset of the parton density distribution.
\item {\p {\p ipar}=\makebox[1.5cm][l]{2017:}
                        \makebox[3cm][l]{({\p EPEE1})}
                        \makebox[3cm][l]{(FLOAT)}
                        \makebox[3cm][l]{(DEF=30.712)}}\\
Defines the laboratory frame by specifying the fraction $E_P/E_e$ in this
frame. It is assumed that the incoming proton is massless.
If the laboratory frame is the rest frame of a fixed target, then the event
record is calculated in the following way \prelim:
At first, the event record is
calculated in a frame given by $E_P/E_e$={\p DABS(EPEE1)}.
This is done under the
assumption that the proton is massless. Then the event record is boosted to
the fixed target frame assuming a massive proton. If the event record
in the fixed target frame is needed,
{\p EPEE1} should be set to some ``large'' negative value,
-10.0 for example.
\item {\p {\p ipar}=\makebox[1.5cm][l]{2018:}
                        \makebox[3cm][l]{({\p DTHEKPMIN})}
                        \makebox[3cm][l]{(FLOAT)}
                        \makebox[3cm][l]{(DEF=0.0)}}\\
Cut on the minimum polar angle of the outgoing lepton
in the laboratory frame (in the range $[0^\circ,180^\circ]$,
$0^\circ$ corresponds to the proton direction).
\item {\p {\p ipar}=\makebox[1.5cm][l]{2019:}
                        \makebox[3cm][l]{({\p DTHEKPMAX})}
                        \makebox[3cm][l]{(FLOAT)}
                        \makebox[3cm][l]{(DEF=180.0)}}\\
Cut on the maximum polar angle of the outgoing lepton
in the laboratory frame (in the range $[0^\circ,180^\circ]$,
$0^\circ$ corresponds to the proton direction).
\item {\p {\p ipar}=\makebox[1.5cm][l]{2020:}
                        \makebox[3cm][l]{({\p DTHEPIMIN})}
                        \makebox[3cm][l]{(FLOAT)}
                        \makebox[3cm][l]{(DEF=0.0)}}\\
Cut on the minimum polar angle of the outgoing partons
in the laboratory frame (in the range $[0^\circ,180^\circ]$,
$0^\circ$ corresponds to the proton direction).
\item {\p {\p ipar}=\makebox[1.5cm][l]{2021:}
                        \makebox[3cm][l]{({\p DTHEPIMAX})}
                        \makebox[3cm][l]{(FLOAT)}
                        \makebox[3cm][l]{(DEF=180.0)}}\\
Cut on the maximum polar angle of the outgoing partons
in the laboratory frame (in the range $[0^\circ,180^\circ]$,
$0^\circ$ corresponds to the proton direction).
\item {\p {\p ipar}=\makebox[1.5cm][l]{2026:}
                        \makebox[3cm][l]{({\p ILAMBDA})}
                        \makebox[3cm][l]{(INT)}
                        \makebox[3cm][l]{(DEF=0)}}\\
\makebox[0.7cm][l]{0:} use $\Lambda_{\mbox{QCD}}$ and
the order of the running coupling
(1-loop, 2-loop) as specified in the parton density parametrization;\\
\makebox[0.7cm][l]{1:} use parameters \#2027, \#2029 instead.
\item {\p {\p ipar}=\makebox[1.5cm][l]{2027:}
                        \makebox[3cm][l]{({\p DLAMBDA})}
                        \makebox[3cm][l]{(FLOAT)}
                        \makebox[3cm][l]{(DEF=0.25)}}\\
$\Lambda_{\mbox{QCD}}$ for 4 flavours, used only if \#2026 is set to 1.
\item {\p {\p ipar}=\makebox[1.5cm][l]{2029:}
                        \makebox[3cm][l]{({\p IARUN})}
                        \makebox[3cm][l]{(int)}
                        \makebox[3cm][l]{(DEF=2)}}\\
Specifies order of running coupling, used only if \#2026 is set to 1.\\
\makebox[0.7cm][l]{1:} 1-loop formula is used;\\
\makebox[0.7cm][l]{2:} 2-loop formula is used.
\item {\p {\p ipar}=\makebox[1.5cm][l]{2030:}
                        \makebox[3cm][l]{({\p XHMI})}
                        \makebox[3cm][l]{(FLOAT)}
                        \makebox[3cm][l]{(DEF=0.0)}}\\
Specifies the lower bound on $x_B$.
\item {\p {\p ipar}=\makebox[1.5cm][l]{2031:}
                        \makebox[3cm][l]{({\p XHMA})}
                        \makebox[3cm][l]{(FLOAT)}
                        \makebox[3cm][l]{(DEF=1.0)}}\\
Specifies the upper bound on $x_B$.
\item {\p {\p ipar}=\makebox[1.5cm][l]{2032:}
                        \makebox[3cm][l]{({\p QQMI})}
                        \makebox[3cm][l]{(FLOAT)}
                        \makebox[3cm][l]{(DEF=0.0)}}\\
Specifies the lower bound on $\sqrt{Q^2}$.
\item {\p {\p ipar}=\makebox[1.5cm][l]{2033:}
                        \makebox[3cm][l]{({\p QQMA})}
                        \makebox[3cm][l]{(FLOAT)}
                        \makebox[3cm][l]{(DEF=$\sqrt{\SH}$)}}\\
Specifies the upper bound on $\sqrt{Q^2}$.
\item {\p {\p ipar}=\makebox[1.5cm][l]{2034:}
                        \makebox[3cm][l]{({\p YMI})}
                        \makebox[3cm][l]{(FLOAT)}
                        \makebox[3cm][l]{(DEF=0)}}\\
Specifies the lower bound on $y$.
\item {\p {\p ipar}=\makebox[1.5cm][l]{2035:}
                        \makebox[3cm][l]{({\p YMA})}
                        \makebox[3cm][l]{(FLOAT)}
                        \makebox[3cm][l]{(DEF=1.0)}}\\
Specifies the upper bound on $y$.
\item {\p {\p ipar}=\makebox[1.5cm][l]{2036:}
                        \makebox[3cm][l]{({\p WWMI})}
                        \makebox[3cm][l]{(FLOAT)}
                        \makebox[3cm][l]{(DEF=0.0)}}\\
Specifies the lower bound on $\sqrt{W^2}$.
\item {\p {\p ipar}=\makebox[1.5cm][l]{2037:}
                        \makebox[3cm][l]{({\p WWMA})}
                        \makebox[3cm][l]{(FLOAT)}
                        \makebox[3cm][l]{(DEF=$\sqrt{\SH}$)}}\\
Specifies the upper bound on $\sqrt{W^2}$.
\item {\p {\p ipar}=\makebox[1.5cm][l]{2038:}
                        \makebox[3cm][l]{({\p ISCALE})}
                        \makebox[3cm][l]{(INT)}
                        \makebox[3cm][l]{(DEF=0)}}\\
Choice of the factorisation scale. This is the scale where the parton densities
are evaluated.\\
\makebox[0.7cm][l]{0:} $\mu_f$=$Q$;\\
\makebox[0.7cm][l]{1:} $\mu_f$={\p RHOSCALE}$\times Q$;\\
\makebox[0.7cm][l]{2:} $\mu_f$={\p RHOSCALE}$\times W$ \prelim;\\
\makebox[0.7cm][l]{3:} $\mu_f$={\p RHOSCALE}$\times \sqrt{\SH}$ \prelim;\\
\makebox[0.7cm][l]{4:} $\mu_f$={\p RHOSCALE}
$\times W^\alpha Q^\beta \sqrt{\SH y}^{1-\alpha-\beta}$ \prelim.
\item {\p {\p ipar}=\makebox[1.5cm][l]{2039:}
                        \makebox[3cm][l]{({\p RHOSCALE})}
                        \makebox[3cm][l]{(FLOAT)}
                        \makebox[3cm][l]{(DEF=1.0)}}\\
Parameter {\p RHOSCALE}
in the choice of the factorisation scale, see \#2038.
\item {\p {\p ipar}=\makebox[1.5cm][l]{2040:}
                        \makebox[3cm][l]{({\p ISCALER})}
                        \makebox[3cm][l]{(INT)}
                        \makebox[3cm][l]{(DEF=0)}}\\
Choice of the renormalisation scale.
This is the scale where the running coupling $\alpha_s$
is evaluated.\\
\makebox[0.7cm][l]{0:} $\mu_r$=$Q$;\\
\makebox[0.7cm][l]{1:} $\mu_r$={\p RHOR}$\times Q$;\\
\makebox[0.7cm][l]{2:} $\mu_r$={\p RHOR}$\times W$ \prelim;\\
\makebox[0.7cm][l]{3:} $\mu_r$={\p RHOR}$\times \sqrt{\SH}$ \prelim;\\
\makebox[0.7cm][l]{4:} $\mu_r$={\p RHOR}
$\times W^\alpha Q^\beta \sqrt{\SH y}^{1-\alpha-\beta}$ \prelim.
\item {\p {\p ipar}=\makebox[1.5cm][l]{2041:}
                        \makebox[3cm][l]{({\p RHOR})}
                        \makebox[3cm][l]{(FLOAT)}
                        \makebox[3cm][l]{(DEF=1.0)}}\\
Parameter {\p RHOR} in the choice of the renormalisation scale, see \#2040.
\item {\p {\p ipar}=\makebox[1.5cm][l]{2042:}
                        \makebox[3cm][l]{({\p DMFMIN})}
                        \makebox[3cm][l]{(FLOAT)}
                        \makebox[3cm][l]{(DEF=0.0)}}\\
\prelim{} Minimal factorization scale (in \GeV).
\item {\p {\p ipar}=\makebox[1.5cm][l]{2043:}
                        \makebox[3cm][l]{({\p DMFMAX})}
                        \makebox[3cm][l]{(FLOAT)}
                        \makebox[3cm][l]{(DEF=$\sqrt{\SH}$)}}\\
\prelim{} Maximal factorization scale (in \GeV).
\item {\p {\p ipar}=\makebox[1.5cm][l]{2044:}
                        \makebox[3cm][l]{({\p DMRMIN})}
                        \makebox[3cm][l]{(FLOAT)}
                        \makebox[3cm][l]{(DEF=0.0)}}\\
\prelim{} Minimal renormalisation scale (in \GeV).
\item {\p {\p ipar}=\makebox[1.5cm][l]{2045:}
                        \makebox[3cm][l]{({\p DMRMAX})}
                        \makebox[3cm][l]{(FLOAT)}
                        \makebox[3cm][l]{(DEF=$\sqrt{\SH}$)}}\\
\prelim{} Maximal renormalisation scale (in \GeV).
\item {\p {\p ipar}=\makebox[1.5cm][l]{2046:}
                        \makebox[3cm][l]{({\p ICUTTYPE})}
                        \makebox[3cm][l]{(INT)}
                        \makebox[3cm][l]{(DEF=0)}}\\
Specifies the invariant mass cut.\\
\makebox[0.7cm][l]{0:} $s_{ij}<cW^2$;\\
\makebox[0.7cm][l]{1:} $s_{ij}<cQ^2$ \prelim;\\
\makebox[0.7cm][l]{2:} $s_{ij}<c\SH$ \prelim;\\
\makebox[0.7cm][l]{3:} $s_{ij}<c
W^\alpha Q^\beta \sqrt{\SH y}^{1-\alpha-\beta}$ \prelim.\\
{\p ICUTTYPE}=1,2,3 is implemented for test purposes only.
\item {\p {\p ipar}=\makebox[1.5cm][l]{2047:}
                        \makebox[3cm][l]{({\p IAEMRUN})}
                        \makebox[3cm][l]{(INT)}
                        \makebox[3cm][l]{(DEF=0)}}\\
Specifies whether the fine structure constant is fixed ($\alpha(0)$)
or running ($\alpha(Q^2)$).\\
\makebox[0.7cm][l]{0:} fixed;\\
\makebox[0.7cm][l]{1:} running.
\item {\p {\p ipar}=\makebox[1.5cm][l]{2048:}
                        \makebox[3cm][l]{({\p PRENA})}
                        \makebox[3cm][l]{(FLOAT)}
                        \makebox[3cm][l]{(DEF=0.5)}}\\
\prelim{} Parameter $\alpha$
used in the calculation of the renormalisation scale.
\item {\p {\p ipar}=\makebox[1.5cm][l]{2049:}
                        \makebox[3cm][l]{({\p PRENB})}
                        \makebox[3cm][l]{(FLOAT)}
                        \makebox[3cm][l]{(DEF=0.5)}}\\
\prelim{} Parameter $\beta$
used in the calculation of the renormalisation scale.
\item {\p {\p ipar}=\makebox[1.5cm][l]{2050:}
                        \makebox[3cm][l]{({\p PFACTA})}
                        \makebox[3cm][l]{(FLOAT)}
                        \makebox[3cm][l]{(DEF=0.5)}}\\
\prelim{} Parameter $\alpha$ used in the calculation of the factorization
scale.
\item {\p {\p ipar}=\makebox[1.5cm][l]{2051:}
                        \makebox[3cm][l]{({\p PFACTB})}
                        \makebox[3cm][l]{(FLOAT)}
                        \makebox[3cm][l]{(DEF=0.5)}}\\
\prelim{} Parameter $\beta$ used in the calculation of the factorization scale.
\item {\p {\p ipar}=\makebox[1.5cm][l]{2052:}
                        \makebox[3cm][l]{({\p PCUTA})}
                        \makebox[3cm][l]{(FLOAT)}
                        \makebox[3cm][l]{(DEF=0.5)}}\\
\prelim{} Parameter $\alpha$ used in the calculation of the invariant mass cut.
\item {\p {\p ipar}=\makebox[1.5cm][l]{2053:}
                        \makebox[3cm][l]{({\p PCUTB})}
                        \makebox[3cm][l]{(FLOAT)}
                        \makebox[3cm][l]{(DEF=0.5)}}\\
\prelim{} Parameter $\beta$ used in the calculation of the invariant mass cut.
\item {\p {\p ipar}=\makebox[1.5cm][l]{2054:}
                        \makebox[3cm][l]{({\p DMCMIN})}
                        \makebox[3cm][l]{(FLOAT)}
                        \makebox[3cm][l]{(DEF=0.0)}}\\
\prelim{} Minimal scale in the invariant mass cut (in \GeV).
\item {\p {\p ipar}=\makebox[1.5cm][l]{2055:}
                        \makebox[3cm][l]{({\p DMCMAX})}
                        \makebox[3cm][l]{(FLOAT)}
                        \makebox[3cm][l]{(DEF=$\sqrt{\SH}$)}}\\
\prelim{} Maximal scale in the invariant mass cut (in \GeV).
\item {\p {\p ipar}=\makebox[1.5cm][l]{2056:}
                        \makebox[3cm][l]{({\p ICLUSTYPE})}
                        \makebox[3cm][l]{(INT)}
                        \makebox[3cm][l]{(DEF=1)}}\\
\makebox[0.7cm][l]{1:} use the mJADE algorithm;\\
\makebox[0.7cm][l]{2:} \prelim{} all partons required to be at $x_F>0$;
cluster outgoing partons with the JADE cluster algorithm,
do {\em not} include the remnant as a precluster.
{\p ICLUSTYPE=2} is implemented for Born terms
only.
\item {\p {\p ipar}=\makebox[1.5cm][l]{2057:}
                        \makebox[3cm][l]{({\p EELMIN})}
                        \makebox[3cm][l]{(FLOAT)}
                        \makebox[3cm][l]{(DEF=0.0D0)}}\\
\prelim{}
Minimal electron energy in the laboratory frame (implemented as a cut).
\item {\p {\p ipar}=\makebox[1.5cm][l]{2058:}
                        \makebox[3cm][l]{({\p PTPVP})}
                        \makebox[3cm][l]{(FLOAT)}
                        \makebox[3cm][l]{(DEF=0.0D0)}}\\
\prelim{} Minimal $p_T$ of all jets (except the remnant jet) in the
proton--virtual--photon CM frame (PVP frame).
\item {\p {\p ipar}=\makebox[1.5cm][l]{2059:}
                        \makebox[3cm][l]{({\p IFPDFLIB})}
                        \makebox[3cm][l]{(INT)}
                        \makebox[3cm][l]{(DEF=0)}}\\
Switch for the parton density library:\\
\makebox[0.7cm][l]{0:} {\p PAKPDF},\\
\makebox[0.7cm][l]{1:} {\p PDFLIB},\\
\makebox[0.7cm][l]{2:} private parametrization \prelim{}.
\item {\p {\p ipar}=\makebox[1.5cm][l]{3002:}
                        \makebox[3cm][l]{({\p NPOINTS})}
                        \makebox[3cm][l]{(INT)}
                        \makebox[3cm][l]{(DEF=100000)}}\\
Number of points in the last {\p VEGAS}
integration. The typical number should be
of the order of 20.000 to 500.000, depending on the complexity of the process
under consideration.
\item {\p {\p ipar}=\makebox[1.5cm][l]{3003:}
                        \makebox[3cm][l]{({\p IALPH})}
                        \makebox[3cm][l]{(INT)}
                        \makebox[3cm][l]{(DEF=1)}}\\
Specifies whether
{\p VEGAS} adjusts the integration grid in every iteration.\\
\makebox[0.7cm][l]{0:} integration grid fixed \prelim;\\
\makebox[0.7cm][l]{1:} adaptive integration.\\
It is recommended to set {\p IALPH}=1.
\item {\p {\p ipar}=\makebox[1.5cm][l]{3004:}
                        \makebox[3cm][l]{({\p NIITMX1})}
                        \makebox[3cm][l]{(INT)}
                        \makebox[3cm][l]{(DEF=10)}}\\
Number of iterations in the {\p VEGAS} grid definition run.
Should be of the order of
10.
\item {\p {\p ipar}=\makebox[1.5cm][l]{3005:}
                        \makebox[3cm][l]{({\p NPOINTS1})}
                        \makebox[3cm][l]{(INT)}
                        \makebox[3cm][l]{(DEF=100000)}}\\
Total number of points in the {\p VEGAS} grid definition run.
Should be of the order
of {\p NPOINTS} to 3*{\p NPOINTS}.
\item {\p {\p ipar}=\makebox[1.5cm][l]{3006:}
                        \makebox[3cm][l]{({\p IPROGRESS})}
                        \makebox[3cm][l]{(INT)}
                        \makebox[3cm][l]{(DEF=1)}}\\
Flag to determine whether the progress of the integration is printed.\\
\makebox[0.7cm][l]{0:} do not print information;\\
\makebox[0.7cm][l]{1:} print information.
\item {\p {\p ipar}=\makebox[1.5cm][l]{3007:}
                        \makebox[3cm][l]{({\p IINTPROGRESS})}
                        \makebox[3cm][l]{(INT)}
                        \makebox[3cm][l]{(DEF=1)}}\\
Flag to determine whether intermediate results
of the adaptive integration routine are printed.\\
\makebox[0.7cm][l]{0:} do not print intermediate results;\\
\makebox[0.7cm][l]{1:} print intermediate results.
\item {\p {\p ipar}=\makebox[1.5cm][l]{4001:}
                        \makebox[3cm][l]{({\p IXSECT(1)})}
                        \makebox[3cm][l]{(INT)}
                        \makebox[3cm][l]{(DEF=1)}}\\
Flag for quark initiated processes\\
\makebox[1.7cm][l]{(1+1) jet:} final state q,\\
\makebox[1.7cm][l]{(2+1) jet:} final state qg,\\
\makebox[1.7cm][l]{(3+1) jet:} final state qgg.\\
\makebox[0.7cm][l]{0:} do not include contribution;\\
\makebox[0.7cm][l]{1:} include contribution.
\item {\p {\p ipar}=\makebox[1.5cm][l]{4002:}
                        \makebox[3cm][l]{({\p IXSECT(2)})}
                        \makebox[3cm][l]{(INT)}
                        \makebox[3cm][l]{(DEF=1)}}\\
Flag for gluon initiated processes ((2+1) and (3+1) jets).\\
\makebox[0.7cm][l]{0:} do not include contribution;\\
\makebox[0.7cm][l]{1:} include contribution.
\item {\p {\p ipar}=\makebox[1.5cm][l]{4003:}
                        \makebox[3cm][l]{({\p IXSECT(3)})}
                        \makebox[3cm][l]{(INT)}
                        \makebox[3cm][l]{(DEF=1)}}\\
Flag for quark initiated process with no gluon in the final state
((3+1) jets only).\\
\makebox[0.7cm][l]{0:} do not include contribution;\\
\makebox[0.7cm][l]{1:} include contribution.
\item {\p {\p ipar}=\makebox[1.5cm][l]{4004:}
                        \makebox[3cm][l]{({\p NPARTONS})}
                        \makebox[3cm][l]{(INT)}
                        \makebox[3cm][l]{(DEF=3)}}\\
Choice of process.\\
\makebox[0.7cm][l]{-3:} (3+1) jets, metric \& longitudinal
polarisations of the virtual photon;\\
\makebox[0.7cm][l]{-2:} (2+1) jets, metric, longitudinal
and other
polarisations of the virtual photon;\\
\makebox[0.7cm][l]{-1:} (1+1) jets, metric \& longitudinal
polarisations of the virtual photon;\\
\makebox[0.7cm][l]{ 1:} (1+1) jets, all polarisations
of the virtual photon;\\
\makebox[0.7cm][l]{ 2:} (2+1) jets, all polarisations
of the virtual photon;\\
\makebox[0.7cm][l]{ 3:} (3+1) jets, all polarisations
of the virtual photon.
\item {\p {\p ipar}=\makebox[1.5cm][l]{4005:}
                        \makebox[3cm][l]{({\p NOUTFLAV})}
                        \makebox[3cm][l]{(INT)}
                        \makebox[3cm][l]{(DEF=5)}}\\
Number of active outgoing quark flavours.
The value of this parameter should depend on $Q^2$ and
on the available phase space
in the hadronic final state.
\item {\p {\p ipar}=\makebox[1.5cm][l]{4006:}
                        \makebox[3cm][l]{({\p I0LOOP})}
                        \makebox[3cm][l]{(INT)}
                        \makebox[3cm][l]{(DEF=1)}}\\
Specifies whether the Born term is included.\\
\makebox[0.7cm][l]{0:} do not include Born term;\\
\makebox[0.7cm][l]{1:} include Born term.\\
0 is not recommended, since the integration tends to be unstable if the Born
term is not included.
\item {\p {\p ipar}=\makebox[1.5cm][l]{4007:}
                        \makebox[3cm][l]{({\p I1LOOP})}
                        \makebox[3cm][l]{(INT)}
                        \makebox[3cm][l]{(DEF=0)}}\\
\prelim $\,$ If a NLO correction is implemented:\\
\makebox[0.7cm][l]{0:} exclude virtual correction;\\
\makebox[0.7cm][l]{1:} include virtual correction.\\
See also parameter \#4012.
\item {\p {\p ipar}=\makebox[1.5cm][l]{4008:}
                        \makebox[3cm][l]{({\p IREALF})}
                        \makebox[3cm][l]{(INT)}
                        \makebox[3cm][l]{(DEF=0)}}\\
\prelim $\,$ If a NLO correction is implemented:\\
\makebox[0.7cm][l]{0:} exclude real final state contributions;\\
\makebox[0.7cm][l]{1:} include real final state contributions.\\
See also parameter \#4012.
\item {\p {\p ipar}=\makebox[1.5cm][l]{4009:}
                        \makebox[3cm][l]{({\p IREALI})}
                        \makebox[3cm][l]{(INT)}
                        \makebox[3cm][l]{(DEF=0)}}\\
\prelim $\,$ If a NLO correction is implemented:\\
\makebox[0.7cm][l]{0:} exclude real initial state contributions;\\
\makebox[0.7cm][l]{1:} include real initial state contributions.\\
See also parameter \#4012.
\item {\p {\p ipar}=\makebox[1.5cm][l]{4011:}
                        \makebox[3cm][l]{({\p I0LOOPLT})}
                        \makebox[3cm][l]{(INT)}
                        \makebox[3cm][l]{(DEF=1)}}\\
In the case of {\p NPARTONS}~$<$~0, this variable
is a binary representation of the polarisations of the virtual photon.\\
\makebox[0.7cm][l]{+1:} metric polarisation;\\
\makebox[0.7cm][l]{+2:} longitudinal polarisation;\\
\makebox[0.7cm][l]{+4:} contribution $\sim \cos \Phi$ ((2+1) jets only);\\
\makebox[0.7cm][l]{+8:} contribution $\sim \cos 2\Phi$ ((2+1) jets only).
\item {\p {\p ipar}=\makebox[1.5cm][l]{4012:}
                        \makebox[3cm][l]{({\p INLO})}
                        \makebox[3cm][l]{(INT)}
                        \makebox[3cm][l]{(DEF=0)}}\\
If a NLO correction is implemented:\\
\makebox[0.7cm][l]{0:} exclude NLO correction;\\
\makebox[0.7cm][l]{1:} include NLO correction.\\
This parameter flag should be used rather than
\#4007, \#4008, \#4009. If this parameter is 0,
the three mentioned parameters are set to 0, if it is 1,
the three parameters are set to 1.
\item {\p {\p ipar}=\makebox[3cm][l]{10012}
                        \makebox[4cm][l]{({\p IDOINT})}
                        \makebox[3cm][l]{(INT)}
                        \makebox[3cm][l]{(DEF=1)}}\\
This flag determines whether the integration is performed ({\p IDOINT=1})
or not ({\p IDOINT=0}). The last option can be used for test purposes.
\item {\p {\p ipar}=\makebox[3cm][l]{11000-11999:}
                        \makebox[6cm][l]{({\p user defined parameters})}
                        \makebox[1cm][l]{()}
                        \makebox[3cm][l]{(no default)}}\\
These flags will not be used in future versions of \PROJP. They may be used for
user defined parameters specified in an input file.
\item {\p {\p ipar}=\makebox[3cm][l]{12001-12100:}
                        \makebox[4cm][l]{({\p IUSERPAR()})}
                        \makebox[3cm][l]{(INT)}
                        \makebox[3cm][l]{(DEF=0)}}\\
These are 100 integer variables for free use. They can be specified in the
input file and are stored in the variable {\p IUSERPAR(i)}, where
{\p ipar=12000+i}.
\item {\p {\p ipar}=\makebox[3cm][l]{12101-12200:}
                        \makebox[4cm][l]{({\p DUSERPAR()})}
                        \makebox[3cm][l]{(FLOAT)}
                        \makebox[3cm][l]{(DEF=0.0)}}\\
These are 100 double precision floating point
variables for free use. They can be specified in the
input file and are stored in the variable {\p DUSERPAR(i)}, where
{\p ipar=12100+i}.
\item {\p {\p ipar}=\makebox[3cm][l]{12201-12300:}
                        \makebox[4cm][l]{({\p CUSERPAR()})}
                        \makebox[3cm][l]{(STRING)}
                        \makebox[3cm][l]{(DEF=' ')}}\\
These are 100 string variables with a length of 100 characters each
for free use. They can be specified in the
input file and are stored in the variable {\p CUSERPAR(i)}, where
{\p ipar=12200+i}. The string is read from the standard input file
({\p unit=5}).
This option is only useful if the parameters are read from an
input file, see below. It is assumed that the string is located
in the beginning of
the next line which is read from the file.
\end{itemize}

\noindent
The common block with user defined variables is {\p /userpar/}:

\begin{verbatim}
      integer iuserint,iuserfloat,iuserstring
      parameter (iuserint=100,iuserfloat=100,iuserstring=100)
      double precision duserpar
      integer iuserpar
      char*100 cuserpar
      common /userpar/
     &       duserpar(iuserfloat),
     &       iuserpar(iuserint),
     &       cuserpar(iuserstring)
\end{verbatim}

\noindent
Table \ref{tab1} shows the the combinations of the parameters
\#4004, \#4006, \#4011 and \#4012 for which there are cross sections
implemented in \PROJP.

\begin{table}[htbp]
\begin{center}
\begin{tabular}{|c|c|c|c|l|}\hline
\#4004 & \#4006 & \#4011 & \#4012 &
\rule[-3mm]{0mm}{8mm} Process\\[0.5ex]\hline\hline
-3&1&+1&0& (3+1), Born, metric\\ \hline
-3&1&+2&0& (3+1), Born, longitudinal\\ \hline
\hline
-2&1&+1&0& (2+1), Born, metric ($\sigma_M$)\\ \hline
-2&1&+2&0& (2+1), Born, longitudinal ($\sigma_L$)\\ \hline
-2&1&+4&0& (2+1), Born, $\sim \cos \Phi$ ($\sigma_{\Phi}$)\\ \hline
-2&1&+8&0& (2+1), Born, $\sim \cos 2\Phi$ ($\sigma_{2\Phi}$)\\ \hline
\hline
-2&1&+1&1& (2+1), Born \& NLO, metric ($\sigma_M$)\\ \hline
-2&1&+2&1& (2+1), Born \& NLO, longitudinal ($\sigma_L$)\\ \hline
-2&1&+4&1& (2+1), Born \& NLO, $\sim \cos \Phi$ ($\sigma_{\Phi}$)\\ \hline
-2&1&+8&1& (2+1), Born \& NLO, $\sim \cos 2\Phi$ ($\sigma_{2\Phi}$)\\ \hline
\hline
-1&1&1&0& (1+1), Born, metric\\ \hline
\hline
-1&1&+1&1& (1+1), Born \& NLO, metric\\ \hline
-1&1&+2&1& (1+1), Born \& NLO, longitudinal\\ \hline
\hline
 3&1&0&0& (3+1), Born, all helicities\\ \hline
\hline
 2&1&0&0& (2+1), Born, all helicities\\ \hline
\hline
 1&1&0&0& (1+1), Born, all helicities\\ \hline
\end{tabular}\\[0.5ex]
\end{center}
\caption{\label{tab1} Cross sections implemented in \PROJP.}
\end{table}

\noindent
In this table, the notation for the parameter \#4011 is the following.
To include a specific helicity, the numbers $i$ indicated by $+i$
have to be added up. To include just the metric and longitudinal contribution
in the case of (2+1) jets, the corresponding parameter value would be
3 (=$1+2$). To include all helicities, the parameter value is 15 ($=1+2+4+8$).
Please note that in the case of (3+1) jets, the helicities that depend
explicitly on the orientation of the leptonic and hadronic systems
can not be selected separately. They are, however, included
in the option ``all helicities''.
In the case of (1+1) jet cross sections, there are no angular dependent
terms, since the orientation of the leptonic and hadronic systems
is specified entirely by the variable $y$.

\noindent
Some remarks concerning the cuts are in order.
The parameters \#2030--\#2037 are implemented by a variable transformation that
transforms the complicated shape given by these cuts into a rectangle. So
imposing these cuts does not lead to an inefficiency in the integration.
The parameters \#2018--\#2021 are
treated differently. The program calculates the
angles of the outgoing particles in the laboratory frame and then simply sets
the differential cross sections to zero if at least one of these cuts is
violated.

\noindent
The application of angular cuts leads
to the following problem: If an acceptance cut of, for
example, $10^\circ$ is specified for the outgoing partons, then, for example,
in the case of (1+1) jet production, only those events are counted as (1+1)
jets which have an outgoing parton with an angle of more than $10^\circ$
relative to the incoming proton. If the jet analysis is done with a cluster
algorithm, there is a second class of events that will be classified as a
(1+1) jet event. Suppose a (2+1) jet event (with all invariants of the outgoing
jets larger than the jet cut) with one of the jets in the direction of
$2^\circ$
relative to the proton, such that this jet disappears in the beam pipe.
This event in principle {\em is} a (2+1) jet event,
but the experimental cluster algorithm would classify this as a (1+1) jet event
because one of the jets is not seen. To get a ``realistic'' (1+1) jet
cross section, one therefore has to add the cross section of {\em real} (2+1)
jet events that {\em look like} (1+1) jet events
to the class of (1+1) jet events.
In order to do this,
the user has to use the event record and
select those events that fulfill the criteria described above.

\noindent
There are several other parameters not appearing in this list defined in
\PROJP.
It is not recommended to use them since they are not necessarily supported in
new versions of the program.

\noindent
Since it is not very flexible to recompile the calling program
every time when a
parameter has been changed, the example program {\p MCEX.F}
demonstrates how to use an input file
{\p PAREX}
to set all parameters.
This input file is a sequential text file.
The file starts with two text strings;
the first gives the filename for an output file, and the second
specifies a job name.
Then there are consecutive lines of the form

\pprog{
   \noindent
   \fulllineend{ipar,par}
}

\noindent
Here {\p ipar} specifies which parameter gets the value {\p par}.
Comments are included in the following way:

\pprog{
   \noindent
   \fulllineend{0,0 This is a comment...}
}

\noindent
Program execution continues after a line

\pprog{
   \noindent
   \fulllineend{-4,0}
}

\noindent
The input file is terminated with a line

\pprog{
   \noindent
   \fulllineend{-5,0}
}

\noindent
which terminates program execution.

\section{Error Conditions}
\label{EC}

In this section a list of the error conditions that might cause a stop
of \PROJ is given.
Wherever possible it has been made sure that the program cannot crash due
to numerical instabilities (if this didn't result in too large a sacrifice of
program efficiency). However, there are several cases in which parameters
do not make sense or are not defined. In these cases, the program will print a
warning or error message on the output file (standard output) and call the {\p
user} error routine.
If the error number is negative, the error is considered to
be fatal, and \PROJS execution will stop subsequently after the {\p user}
subroutine for program termination has been called. If the error number is
positive, it is just a warning, and \PROJS continues after making a reasonable
adjustment of the condition that caused the error.
In several cases, the warning messages and the calls to {\p user} are done only
a certain number of times to avoid a messy output. The following is a list
of the errors that might occur:

\begin{itemize}
\item {\p ierr = }\makebox[1.5cm][l]{\p +1:}\\
$Q^2$ not valid in parton density
parametrization, use maximum/minimum $Q^2$ as specified in the particular
parametrization.
\item {\p ierr = }\makebox[1.5cm][l]{\p +2:}\\
An error has been reported from {\p userset}.
\item {\p ierr = }\makebox[1.5cm][l]{\p +3:}\\
Value of {\p KPAR} not valid.
\item {\p ierr = }\makebox[1.5cm][l]{\p -6:}\\
Momentum fraction $\xi$ of the
incoming parton too low, parton densities no longer valid.
\item {\p ierr = }\makebox[1.5cm][l]{\p -8:}\\
Form of running $\alpha_s$ not
known.
\item {\p ierr = }\makebox[1.5cm][l]{\p -9:}\\
$\mu_f^2$ too low in running
$\alpha_s$.
\item {\p ierr = }\makebox[1.5cm][l]{\p +10:}\\
Lepton variable $y$ out of range.
\item {\p ierr = }\makebox[1.5cm][l]{\p -12:}\\
Invariants for virtual (2+1) jet
correction out of range.
\item {\p ierr = }\makebox[1.5cm][l]{\p -13:}\\
Error in a call to the parton
density parametrization.
\item {\p ierr = }\makebox[1.5cm][l]{\p -14:}\\
$Q^2$, $W^2$ bounds not compatible
with $x_B$, $y$ bounds.
\item {\p ierr = }\makebox[1.5cm][l]{\p -15:}\\
No space in the event record.
\item {\p ierr = }\makebox[1.5cm][l]{\p +17:}\\
{\p dacos} function not defined,
argument corrected.
\item {\p ierr = }\makebox[1.5cm][l]{\p +18:}\\
{\p dsqrt} function not defined,
argument corrected.
\item {\p ierr = }\makebox[1.5cm][l]{\p +19:}\\
{\p dlog} function not defined,
argument corrected.
\item {\p ierr = }\makebox[1.5cm][l]{\p -20:}\\
{\p igrid} not valid.
\item {\p ierr = }\makebox[1.5cm][l]{\p -21, -22:}\\
Error when {\p PROJET} tried to access parton density functions.
\end{itemize}

\noindent
Error handling in \PROJS needs some
improvement, since it is not desirable that the program stops execution
completely in case of a fatal error. This will be accomplished step by step in
new versions of the program.

\section{Installation}

The \PROJS package consists of
\begin{itemize}
\item \makebox[4.0cm][l]{{\tt PROJET.TEX}:}
the {\LaTeX} file of the manual;
\item \makebox[4.0cm][l]{{\tt PROJET.F}:}
a set of subroutines;
\item \makebox[4.0cm][l]{{\tt PROJET.MCEX.F}:}
an example program;
\item \makebox[4.0cm][l]{{\tt PROJET.PAREX}:}
an input file for the example program;
\item \makebox[4.0cm][l]{{\tt PROJET.RESEX}:}
the output file ({\p unit=6}) that
{\p MCEX} creates;
\item \makebox[4.0cm][l]{{\tt PROJET.PLOTEX}:}
the output on an external file ({\p unit=55}) that
{\p MCEX} creates;
\item \makebox[4.0cm][l]{{\tt PROJET.PAKPDF.F}:} the library of
parton density parametrizations {\p PAKPDF} \cite{Cha92};
\item \makebox[4.0cm][l]{{\tt PROJET.PAKMAN}:} the {\p PAKPDF} manual;
\item \makebox[4.0cm][l]{{\tt PROJET.PDFDUMMY.F}:}
a dummy routine to
replace the {\p PDFLIB} library of parton density
par\-ametrizations (if not available).
\end{itemize}

\noindent
{\tt PROJET.F} is the set of subroutines containing the cross section
formulas. The sets of subroutines {\tt PROJET.F}, {\tt PROJET.PAKPDF.F},
{\tt PROJET.PDFDUMMY.F} and a main program
have to be linked in order to get an executable program.
A simple example for a main program is {\tt PROJET.MCEX.F}.

\begin{sloppypar}
\noindent
In order to use the {\p PDFLIB} \cite{Plo93}
library of parton density functions,
{\tt PROJET.PDFDUMMY.F} has to be omitted when linking \PROJP.
Parameter \#2059 is the flag that determines the choice of the
parton density library.
\end{sloppypar}

\noindent
The files are available on the DESY IBM mainframe
({\p I02GAU.PROJETxx.yyy})
or by electronic mail on request.
In the near future, {\p PROJET} will be available via {\p FreeHEP}
on the World Wide Web as well.

\section{Summary and Conclusions}
\label{SaC}

\PROJS is a parton level Monte Carlo program for the calculation of jet cross
sections in deeply inelastic electron proton scattering. In the present
version,
the matrix elements for (1+1), (2+1) and (3+1)
jet Born terms and
(1+1) and (2+1) jet NLO terms for
all polarisations of the exchanged virtual photon
are implemented. An event record for every event is accessible by the user.
This permits
the calculation of event variables as well as the implementation
of phase space cuts on the outgoing jets and the outgoing electron.

\noindent
Please send questions, comments and suggestions to graudenz~@~cernvm.cern.ch.
If you want to receive updated versions of this manual,
please send me your electronic mail
address.
I would like to
encourage users to report problems with the program
including a description of the problem, the parameter file
and some remarks on the environment of the program (machine, operating system,
{\p FORTRAN} compiler)
to me.

\noindent
Up to now, \PROJS has been tested on several workstations (including
HP, DEC and SUN workstations) and on a VAX cluster.
It has not been checked whether there are
any incompatibilities with {\p FORTRAN} on IBM mainframes.

\section{Acknowledgements}
\label{ack}

I would like to thank Ch.~Berger, H.~He{\ss}ling, G.~Ingelman,
G.~Kramer, N.~Magnussen, C.~Salgado and
H.~Spiesberger
for interesting and stimulating discussions. I am grateful in particular to
R.~Nisius for discussions, his efforts to test the program
and helpful suggestions for the adaptation of the program to the needs
of experimental physicists.
H. Spiesberger has provided me with the {\p FORTRAN} code for
the calculation of
the scale dependent fine structure constant.
Support from the computer centers of CERN, DESY, LBL and the RWTH Aachen is
gratefully acknowledged.
This work was partly supported by a grant from the Max Kade Foundation.

\newpage
%
%
\newcommand{\bibitema}[1]{\bibitem[#1]{#1}}

\newpage
\begin{appendix}
\section{Test Run}

This, appendix contains the source code of the test run program {\p MCEX.F},
the input file {\p PAREX}, and the output files {\p RESEX} and {\p PLOTEX}.

\subsection{Test Run Program}
\begin{small}
\begin{verbatim}
C ---------------------------------------------------------------------
C
C ---> JET MONTE CARLO PROGRAM FOR DIS
C      BASED ON PROJET
C
C ---> DIRK GRAUDENZ
C
C ---> SOURCE FILE: MCEX.F
C
C ---> 15\04\1993
C      27\08\1994
C
C ---------------------------------------------------------------------

      PROGRAM MCEX

      IMPLICIT DOUBLE PRECISION (A-H,O-Z)

      CHARACTER*32 JOBN,PLOTN
      COMMON /SVAR/
     &       JOBN,PLOTN

C --- INITIALIZATION
      CALL SETPAR(-3,0.D0,IERR)
      IF (IERR .NE. 0) CALL ERRMSG(IERR)
      CALL SETPAR(-2,0.D0,IERR)
      IF (IERR .NE. 0) CALL ERRMSG(IERR)

C --- READ FILENAMES
      READ(5,*)  PLOTN
      READ(5,*)  JOBN
      WRITE(6,*) 'OUTPUT FILE:',PLOTN
      WRITE(6,*) 'JOBNAME    :',JOBN

C --- OPEN OUTPUT FILE
      OPEN(UNIT   =55,
     &     FILE   =PLOTN,
     &     ACCESS ='SEQUENTIAL',
     &     FORM   ='FORMATTED',
     &     STATUS ='UNKNOWN')

C --- READ PARAMETERS FROM EXTERNAL FILE
101   CONTINUE
         READ(5,*) ICARD,XDATA
         IDATA=IDNINT(XDATA)
         IF (ICARD .EQ. -5) GOTO 99
         IF (ICARD .EQ. -4) GOTO 102
         CALL SETPAR(ICARD,XDATA,IERR)
         IF (IERR .NE. 0) CALL ERRMSG(IERR)
      GOTO 101
 102  CONTINUE

C ------ DEFINE PROCESSES AND START INTEGRATIONS

C ------ 1+1, BORN
         CALL SETP(1,1,0,0)

C ------ 1+1, BORN & NLO, TRANSVERSE
         CALL SETP(-1,1,1,1)

C ------ 2+1, BORN
         CALL SETP(2,1,0,0)

C ------ 3+1, BORN
         CALL SETP(3,1,0,0)

C ------ 2+1, BORN, TRANSVERSE
         CALL SETP(-2,1,1,0)

C ------ 2+1, BORN & NLO, TRANSVERSE
         CALL SETP(-2,1,1,1)

C --- CLOSE OUTPUT FILE
      CLOSE(55)

99    CONTINUE
      STOP
      END

C ---------------------------------------------------------------------
C
C === SET PARAMETERS #4004, #4006, #4011, #4012
C     AND START INTEGRATION
C
C ---------------------------------------------------------------------

      SUBROUTINE SETP(I4004,I4006,I4011,I4012)

      CALL SETPAR(4004,DFLOAT(I4004),IERR)
      CALL SETPAR(4006,DFLOAT(I4006),IERR)
      CALL SETPAR(4011,DFLOAT(I4011),IERR)
      CALL SETPAR(4012,DFLOAT(I4012),IERR)
      CALL SETPAR(-1  ,DFLOAT(    0),IERR)

      RETURN
      END

C ---------------------------------------------------------------------
C
C === USER SPECIFIC SUBROUTINES
C
C ---------------------------------------------------------------------

      SUBROUTINE USER(IWHAT,IUERR1)

      IMPLICIT DOUBLE PRECISION (A-H,O-Z)

      INTEGER NMAXENT
      PARAMETER (NMAXENT=20)
      DOUBLE PRECISION POUT,PPOL
      INTEGER NENTRY,IPTYPE,IIDENT,INOUT
      COMMON /EVTRECORD/
     &       POUT(NMAXENT,4),
     &       PPOL(NMAXENT,4),
     &       NENTRY,IPTYPE(NMAXENT),IIDENT(NMAXENT),INOUT(NMAXENT)

      COMMON /PVPRECORD/
     &       PVP(NMAXENT,4),
     &       PVPPOL(NMAXENT,4)

      DOUBLE PRECISION EVWGT,EVXSECT,EVERROR
      INTEGER IEVACCPT
      COMMON /USERWGT/
     &       EVWGT,
     &       EVXSECT,EVERROR,
     &       IEVACCPT

      INTEGER IUSERINT,IUSERFLOAT,IUSERSTRING
      PARAMETER (IUSERINT=100,IUSERFLOAT=100,IUSERSTRING=100)
      DOUBLE PRECISION DUSERPAR
      INTEGER IUSERPAR
      CHARACTER*100 CUSERPAR
      COMMON /USERPAR/
     &       DUSERPAR(IUSERFLOAT),
     &       IUSERPAR(IUSERINT),
     &       CUSERPAR(IUSERSTRING)

      COMMON /MEMO/
     &       WGTSUM,
     &       IUERR

      IF (IWHAT .EQ. 1) THEN
C ------ PRESET
      ELSEIF (IWHAT .EQ. 2) THEN
C ------ INITIALIZATION
      ELSEIF (IWHAT .EQ. 3) THEN
C ------ BEFORE START OF INTEGRATION
         IUERR=0
         WGTSUM=0.
      ELSEIF (IWHAT .EQ. 4) THEN
         IF (IEVACCPT .EQ. 1) THEN
            WGTSUM=WGTSUM+EVWGT
         ENDIF
      ELSEIF (IWHAT .EQ. 5) THEN
C ------ FINISH
         IF (IUERR .NE. 0) THEN
            WRITE(55,*) 'CRASHED! IUERR=',IUERR
         ELSE
C --------- RESULT
            WRITE(6,*) '## INTEGRATED CROSS SECTION=', EVXSECT
            WRITE(6,*) '## ESTIMATED ERROR         =', EVERROR
            WRITE(55,*) 'XSECT =',WGTSUM
         ENDIF
      ELSEIF (IWHAT .EQ. 6) THEN
C ------ ERROR
         IF (IUERR .EQ. 0 .AND. IUERR1 .LT. 0) THEN
            IUERR=IUERR1
         ENDIF
      ENDIF

      RETURN
      END

C ---------------------------------------------------------------------

      SUBROUTINE USERSET(IPAR,IDATA,XDATA,IERR)

      IMPLICIT DOUBLE PRECISION (A-H,O-Z)

      IF (IPAR .EQ. -3) THEN
         IERR=0
         GOTO 99
      ENDIF

C --- IF CALLED, THE PARAMETER IS NOT KNOWN TO PROJET,
C     AND HENCE NOT TO MCEX.
      IERR=-1

99    CONTINUE
      RETURN
      END

C ---------------------------------------------------------------------
C
C === USER SUPPLIED CUTS
C
C ---------------------------------------------------------------------

      FUNCTION IUSERCUT()

      IMPLICIT DOUBLE PRECISION (A-H,O-Z)

C --- NO ADDITIONAL CUTS
      IUSERCUT=1

99    CONTINUE
      RETURN
      END

C ---------------------------------------------------------------------
\end{verbatim}
\end{small}

\newpage
\subsection{Test Run Input}
\begin{small}
\begin{verbatim}
'plotex'
'jobex'
0,0 ====================================================================
0,0     PARAMETER FILE FOR PROJET
0,0     MAIN PROGRAM IS MCEX.F
0,0 ====================================================================
0,0 --- MODE 4: INTEGRATE OVER XB AND Q**2 WITHIN BOUNDS
2001, 4
0,0 --- DO NOT PRINT PROGRESS INFORMATION
3006, 0
3007, 0
0,0 --------------------------------------------------------------------
0,0     KINEMATICS AND CUTS
0,0 --------------------------------------------------------------------
0,0 --- CM-ENERGY
2002, 295.93
0,0 --- TYPE OF CUT (W**2 (0), Q**2 (1), SH (2))
2046, 0
0,0 --- INVARIANT MASS CUT
2008, 0.02
0,0 --- CUT ON MOMENTUM FRACTION CARRIED BY THE INCOMING PARTON
2014, 1.D-5
0,0 --- PARAMETRIZATION OF THE PARTON DENSITIES (SET/PARAM.)
2015, 6
2016, 2
0,0 --- FRACTION OF PROTON ENERGY AND ELECTRON ENERGY IN THE LAB FRAME
2017, 30.712
0,0 --- MINIMUM, MAXIMUM XH
2030, 0.0
2031, 1.0
0,0 --- MINIMUM, MAXIMUM Q
2032, 2.5
2033, 20
0,0 --- MINIMUM, MAXIMUM Y
2034, 0.0
2035, 0.5
0,0 --- MINIMUM, MAXIMUM W
2036, 0.0
2037, 50.0
0,0 ====================================================================
0,0     VEGAS-PARAMETERS
0,0 --------------------------------------------------------------------
0,0 --- NUMBER OF ITERATIONS IN THE GRID-DEFINING RUN
3004, 10
0,0 --- NUMBER OF POINTS IN THE GRID-DEFINING RUN
3005, 20000
0,0 --- NUMBER OF POINTS IN THE MAIN RUN
3002, 20000
0,0 ====================================================================
0,0 --- NUMBER OF ACTIVE OUTGOING FLAVOURS
4005, 5
0,0 ====================================================================
0,0 --- CONTINUE WITH PROGRAM EXECUTION (I.E. PERFORM INTEGRATION)
-4,0
0,0 --- TERMINATE PROGRAM
-5,0
0,0 ====================================================================
\end{verbatim}
\end{small}

\newpage
\subsection{Test Run Output, Part 1}
\begin{small}
\begin{verbatim}



        _/_/_/_/    _/_/_/_/     _/_/_/    _/_/_/_/  _/_/_/_/ _/_/_/_/_/
       _/      _/  _/      _/  _/     _/        _/  _/           _/
      _/      _/  _/      _/  _/     _/        _/  _/           _/
     _/_/_/_/    _/_/_/_/    _/     _/        _/  _/_/_/_/     _/
    _/          _/      _/  _/     _/        _/  _/           _/
   _/          _/       _/ _/     _/  _/    _/  _/           _/
  _/          _/       _/   _/_/_/    _/_/_/   _/_/_/_/     _/



                         |
    ... THE MONTE CARLO PROGRAM FOR
                        JET CROSS SECTIONS
                         |        IN DEEPLY INELASTIC SCATTERING



       +++++++++++++++++++++++++++++++++++++++++++++++++++
       +                                                 +
       +                P R O J E T  4.1                 +
       +                                                 +
       +                AUGUST 28,  1994                 +
       +                                                 +
       +                                                 +
       +  QUESTIONS, COMMENTS AND SUGGESTIONS TO         +
       +  D. GRAUDENZ, GRAUDENZ @ CERNVM.CERN.CH         +
       +  THEORETICAL PHYSICS DIVISION, CERN             +
       +  PHONE: +41-22-767-2959, FAX: +41-22-782-3914   +
       +                                                 +
       +++++++++++++++++++++++++++++++++++++++++++++++++++



OUTPUT FILE:D/plotex
JOBNAME    :ex
 DATA CARD:        2001    0.400000D+01           4
 DATA CARD:        3006    0.000000D+00           0
 DATA CARD:        3007    0.000000D+00           0
 DATA CARD:        2002    0.295930D+03         296
 DATA CARD:        2046    0.000000D+00           0
 DATA CARD:        2008    0.200000D-01           0
 DATA CARD:        2014    0.100000D-04           0
 DATA CARD:        2015    0.600000D+01           6
 DATA CARD:        2016    0.200000D+01           2
 DATA CARD:        2017    0.307120D+02          31
 DATA CARD:        2030    0.000000D+00           0
 DATA CARD:        2031    0.100000D+01           1
 DATA CARD:        2032    0.250000D+01           3
 DATA CARD:        2033    0.200000D+02          20
 DATA CARD:        2034    0.000000D+00           0
 DATA CARD:        2035    0.500000D+00           1
 DATA CARD:        2036    0.000000D+00           0
 DATA CARD:        2037    0.500000D+02          50
 DATA CARD:        3004    0.100000D+02          10
 DATA CARD:        3005    0.200000D+05       20000
 DATA CARD:        3002    0.200000D+05       20000
 DATA CARD:        4005    0.500000D+01           5
 DATA CARD:        4004    0.100000D+01           1
 DATA CARD:        4006    0.100000D+01           1
 DATA CARD:        4011    0.000000D+00           0
 DATA CARD:        4012    0.000000D+00           0

 TOTAL ## OF POINTS        :     19602
 ## SURVIVING THE CUTS     :     19602
 REQUESTED ## OF PTS/IT    :     20000

 VEGAS-INTEGRAL  =                0.649022D+05
 EST. ERROR ABS. =                0.156665D+02
 EST. ERROR IN % =                  0.02413857

## INTEGRATED CROSS SECTION=    64902.203125000
## ESTIMATED ERROR         =    15.666465759277
 DATA CARD:        4004   -0.100000D+01          -1
 DATA CARD:        4006    0.100000D+01           1
 DATA CARD:        4011    0.100000D+01           1
 DATA CARD:        4012    0.100000D+01           1

 TOTAL ## OF POINTS        :     18522
 ## SURVIVING THE CUTS     :     18522
 REQUESTED ## OF PTS/IT    :     20000

 VEGAS-INTEGRAL  =                0.284903D+05
 EST. ERROR ABS. =                0.109468D+03
 EST. ERROR IN % =                  0.38423055

## INTEGRATED CROSS SECTION=    28490.269531250
## ESTIMATED ERROR         =    109.46831512451
 DATA CARD:        4004    0.200000D+01           2
 DATA CARD:        4006    0.100000D+01           1
 DATA CARD:        4011    0.000000D+00           0
 DATA CARD:        4012    0.000000D+00           0

 TOTAL ## OF POINTS        :     15552
 ## SURVIVING THE CUTS     :     14117
 REQUESTED ## OF PTS/IT    :     20000

 VEGAS-INTEGRAL  =                0.247556D+05
 EST. ERROR ABS. =                0.119762D+03
 EST. ERROR IN % =                  0.48377723

## INTEGRATED CROSS SECTION=    24755.607421875
## ESTIMATED ERROR         =    119.76199340820
 DATA CARD:        4004    0.300000D+01           3
 DATA CARD:        4006    0.100000D+01           1
 DATA CARD:        4011    0.000000D+00           0
 DATA CARD:        4012    0.000000D+00           0

 TOTAL ## OF POINTS        :     19683
 ## SURVIVING THE CUTS     :      9632
 REQUESTED ## OF PTS/IT    :     20000

 VEGAS-INTEGRAL  =                0.209663D+04
 EST. ERROR ABS. =                0.365540D+03
 EST. ERROR IN % =                 17.43468094

## INTEGRATED CROSS SECTION=    2096.6281738281
## ESTIMATED ERROR         =    365.54043579102
 DATA CARD:        4004   -0.200000D+01          -2
 DATA CARD:        4006    0.100000D+01           1
 DATA CARD:        4011    0.100000D+01           1
 DATA CARD:        4012    0.000000D+00           0

 TOTAL ## OF POINTS        :     19683
 ## SURVIVING THE CUTS     :     17623
 REQUESTED ## OF PTS/IT    :     20000

 VEGAS-INTEGRAL  =                0.191440D+05
 EST. ERROR ABS. =                0.111850D+03
 EST. ERROR IN % =                  0.58425546

## INTEGRATED CROSS SECTION=    19143.990234375
## ESTIMATED ERROR         =    111.84981536865
 DATA CARD:        4004   -0.200000D+01          -2
 DATA CARD:        4006    0.100000D+01           1
 DATA CARD:        4011    0.100000D+01           1
 DATA CARD:        4012    0.100000D+01           1

 TOTAL ## OF POINTS        :     19683
 ## SURVIVING THE CUTS     :     17849
 REQUESTED ## OF PTS/IT    :     20000

 VEGAS-INTEGRAL  =                0.209589D+05
 EST. ERROR ABS. =                0.207645D+03
 EST. ERROR IN % =                  0.99072427

## INTEGRATED CROSS SECTION=    20958.925781250
## ESTIMATED ERROR         =    207.64517211914
\end{verbatim}
\end{small}

\subsection{Test Run Output, Part2}
\begin{small}
\begin{verbatim}
XSECT =    64902.196744796
XSECT =    28490.199754252
XSECT =    24755.394848660
XSECT =    2096.5009340818
XSECT =    19143.702981131
XSECT =    20958.925678212
\end{verbatim}
\end{small}

\end{appendix}

\end{document}